\documentstyle[aps]{revtex}
\begin{document}
 \draft

\twocolumn[\hsize\textwidth\columnwidth\hsize\csname
@twocolumnfalse\endcsname
\preprint{SU-ITP-98-60\\ hep-th/9811141\\ November 1998}
\date{15 November 1998}
\title{Inflation and Large Internal Dimensions}
\author{Nemanja Kaloper and Andrei Linde
 }
\address{
Department of Physics, Stanford University, Stanford,
CA 94305-4060, USA  }
\maketitle
\begin{abstract}
We consider some aspects of inflation in models with large internal
dimensions. If inflation occurs
on a $3D$ wall after the stabilization of internal dimensions in the
models with low unification scale ($M \sim 1$ TeV ), the inflaton
field must be extremely light. This problem may disappear  In models
with intermediate ($M \sim 10^{11}$ GeV) to high ($M \sim 10^{16}$
GeV) unification scale.  However, in all of these cases the wall
inflation does not provide a complete solution to the horizon and
flatness problems. To solve them, there must be a stage of inflation
in the bulk before the compactification of internal dimensions.
\end{abstract}
\pacs{PACS: 98.80.Cq\hskip 3.6cm SU-ITP-98-60\hskip 3.6cm hep-th/9811141}
\vskip2pc]

The hierarchy problem has been one of the long-standing
challenges to theoretical physics. It is a puzzle concerning
masses of scalar fields, which are all quadratically
divergent in the loop expansion of a
generic quantum field theory. Since the
natural cut-off of any quantum field theory
is the Planck mass $M_p$, the renormalization effects should
drive all scalar masses up to the Planck scale. As a result,
all interacting scalar fields should be very heavy.
On the other hand, in the Standard Model the Higgs field must
have mass $m_H \sim 1$ TeV  in order for the model to be consistent.
This is the hierarchy problem: the light scalars are
needed in the theory, but ensuring that they are light
requires not only that their masses are very small
classically, but also that there is some mechanism which will
keep the masses small after radiative corrections.

However, if one starts with a fundamental theory
which is higher-dimensional,
and identifies the higher-dimensional fundamental
Planck scale with the gauge unification scale $M$,
it may be possible to recover the very large Planck scale of the
four-dimensional ($4D$) world if the higher-dimensional
theory is compactified to $4D$ on a large
internal space
\cite{savas}-\cite{randal}.
There have also been previous considerations of models
of unification scales below the $4D$ Planck scale
in string theory \cite{witt}-\cite{biq}.
If the size of the internal dimensions is $r_0$,
using Newton's law in $D+4$ dimensions, at distances much larger than
the size $r_0$ of the internal space, one finds \cite{savas}
\begin{equation}\label{masses}
M_p^2 \sim r_0^D M^{2+D}
\end{equation}
Thus if $r_0 \gg M^{-1}$, the reduced Planck mass may appear to be
many orders of magnitude larger than the fundamental Planck mass.
The hierarchy problem then becomes the problem of choosing
the radius of stabilization of the internal space, which
should be large compared to the fundamental scale.
In all of these models, it has been assumed that the $4D$ world is
a three-dimensional analogue of a
domain wall, or a $3$-brane in the modern $M$-theory parlance,
which is embedded in a higher-dimensional theory.
The proposal that the world may be a hypersurface in a
higher-dimensional spacetime goes back to
\cite{rs} (see also \cite{ds}), but has
been reinvigorated by recent developments
in string and $M$-theory, which may provide the mechanism to explain
why matter degrees of freedom are stuck to
the wall. In this paper we will give some comments concerning the
possibility to have inflation in such a scenario.

Generically, inflation may begin within a small island of $D+4$
dimensional space of Planck size $M^{-1}$. Then it may proceed
differently in $3$ uncompactified directions and in the remaining $D$
dimensions, which grow from $M^{-1}$ to $r_0$ and then stabilize.
Unfortunately, it is very difficult to study this possibility
since many aspects of compactification and stabilization of $D$
dimensions in this theory still remain rather speculative.
Therefore prior to the investigation of this generic but complicated
regime, one may try to analyze a simpler possibility, assuming that
inflation (or at least its latest stages) occur only in $3$ uncompactified
directions after the stabilization of internal dimensions.

In \cite{bendav,lyth}
it has been argued that having all of inflation
occur after compactification may require an extremely light wall
inflaton, as compared to the unification scale.
Indeed, the effective potential along the wall in this scenario cannot be
greater
than $M^4$. This follows from the assumption that the thickness of
the wall cannot be much greater than $M^{-1}$, and the Planck density in
$D+4$ is $M^{D+4}$. Then the Hubble constant during inflation on the wall
is given by
\begin{equation}
\label{2}
H  \sim \sqrt{8\pi V(\phi)\over 3 M_p^2} \lesssim {M^2\over  M_p} \ .
\end{equation}

Inflation occurs only if the inflaton mass $m$ is smaller than $H$, which
implies the constraint
\begin{equation}\label{inflatonmass}
m \lesssim H   \lesssim {M^2\over  M_p} \ .
\end{equation}
This bound is completely independent of the number
of internal dimensions. In the particular case $D = 2$, this constraint
shows that the Compton wavelength of the inflaton field should be greater
than the size of internal dimensions.

If one takes $M \sim 1$ TeV, as proposed in \cite{savas}, one gets
an extremely
strong constraint on the inflaton mass,
$m \lesssim  {10^{-4}} {\mbox eV}$ \cite{bendav,lyth}.
In principle, supersymmetry may provide some flat directions with an
extremely
small curvature $V''= m^2 < ({10^{-4}} {\mbox eV})^2$, but this
forces one to
make a step back from the original motivation for the models of this type.
Even if this is allowed, one still encounters
severe problems in constructing inflationary models of such type.

For example, if one considers a simplest version of chaotic inflation with
$V(\phi) = {m^2\over 2}\phi^2$ \cite{andc},
one finds unacceptably density perturbations
\begin{equation}\label{5}
{\delta\rho\over \rho} \sim 50 {m\over M_p}  \lesssim 10^{-31} \ .
\end{equation}

The situation becomes slightly better  for hybrid inflation driven by the
potential
$V(\phi, \sigma) = \frac1{4V }
(M^2 - \lambda  \sigma^2)^2 +
\frac{m^2}{2} \phi^2 +
\frac{g^2}{2} \phi^2 \sigma^2$ \cite{andh,term}. In this case
\begin{equation}
\frac{\delta \rho}{\rho} = \frac{2\sqrt{6\pi}g}{5\lambda ^{3/2}}
\frac{M^5}{M^3_p m^2} \ .
\label{drrhybrid}
\end{equation}
Let us take  $M  \sim 1$ TeV  and $\lambda , g \sim O(1)$, as should be
if the inflaton is to be a particle
from the spectrum of the wall gauge theory.
Compared to the COBE data, which give
$\frac{\delta \rho}{\rho} \sim 5\times 10^{-5}$
at redshifts corresponding to the
last $60$ efoldings of inflation,
we find the desirable value of $m$:
\begin{equation}\label{4big}
m \sim  {10^{-10}} ~{\mbox eV} \ .
\end{equation}
This is 6 orders of magnitude worse than the constraint $m \lesssim
{10^{-4}} {\mbox eV}$ obtained in \cite{bendav,lyth}
from the condition of
existence of the inflationary regime (\ref{inflatonmass}). A different
possibility was discussed in \cite{lyth}, but it requires the
existence of a small coupling constant $\lambda  \sim 10^{-8}$.

Note, that in the hybrid inflation scenario discussed above, the mass
of the inflaton field $\phi$ {\it after } inflation is equal to
$g M/\sqrt \lambda  \gg m$. Therefore one can have efficient
reheating and baryogenesis in this model. Thus, it is possible to
have a consistent inflationary scenario of this type if one finds a
mechanism which maintains the extreme flatness of the effective potential
during inflation. Supersymmetry may help here, but typically supersymmetry
induces the inflaton mass $m = O(H)$. There exist several mechanisms
which may help to avoid this complication \cite{Dterm}.
However, according to Eq. (\ref{4big}), in the model described above
the mass of the inflaton field during inflation must be six orders of
magnitude smaller than $H$, which may be rather difficult to achieve.

The constraint on the inflaton mass can be relaxed by assuming that the
scale $M $ is much larger than $1$~TeV. For example, if we take
$M  \sim 10^{11} GeV$, as suggested in \cite{benakli,biq},
the constraint on the inflaton mass is
\begin{equation}
m < {M^2/M_p} \sim  1 ~ {\mbox TeV}.
\label{hiinfl}
\end{equation}
It fits perfectly in the hybrid inflation scenario. Indeed, in the
original version of the hybrid inflation model \cite{andh} it has been
proposed to take the parameters $M= 10^{11}$ GeV, $m = 10^2$ GeV,
$g^2 = \lambda  = 0.1$, which satisfy the constraint $m < {M^2\over M_p}$
and give  the proper amplitude of density perturbations.

Another interesting possibility is if the unification
scale is $M = 10^{16} - 10^{17}$ GeV  \cite{witt}.
This would lead to the constraint
\begin{equation}
m  \le 10^{13} - 10^{15} ~{\mbox GeV}.
\label{hiinfla}
\end{equation}
This condition is satisfied in the simplest version of chaotic inflation
scenario with $V(\phi) = {m^2\over 2} \phi^2$ and  $m \sim 10^{13}$
GeV \cite{book}.  Hybrid inflation works in this case as well, for a
smaller
value of $m$ \cite{Dterm}.

The discussion above shows that having inflation on the wall
when the unification scale is low, $\sim 1$ TeV, requires incredibly small
masses and couplings. On the other hand, it is rather easy to fit
some of the mainstream inflationary models on the wall in
theories with large internal dimensions which have stabilized, if the
unification scale is medium to high.

However we must point out that having all of inflation after the
internal dimensions are stabilized cannot be the whole
picture. For very low unification scales, the compactification scale
$r^{-1}_0$ and the unification scale
$M $ differ by many orders of magnitude. In such models,
inflation after stabilization of the internal dimensions requires extreme
fine tuning of the initial conditions in the early universe.

Indeed, the only natural timescale for the beginning of inflation in this
model is given by the higher-dimensional Planck time $M^{-1}$, when the
density of the universe was of the order $M^{D+4}$. The last condition  is
consistent with the requirement that the 4D density of the wall is
$O(M^4)$.
However, as we already mentioned, the Hubble parameter $H$ at this time is
smaller than $M^2/M_p  \ll M $, which implies that inflation occurs
on a time
scale much greater than $M^{-1}$. Thus the universe must be sufficiently
large and homogeneous from the very beginning to survive and not to
loose the
homogeneity during the long period of time from $t \sim M^{-1}$ to
$t \sim H^
{-1}$. (If $M \sim 1$ TeV and $H \sim 10^{-4}$ eV, these two time scales
differ by 16 orders of magnitude.) On the other hand, one cannot simply
assume that the universe must be homogeneous at all times, because at the
beginning of  inflation it must be strongly inhomogeneous: its
density at the
wall must be many orders of magnitude greater than the density in
the bulk.
Indeed,  suppose  for definiteness that the initial value of $H$ during
inflation is equal to $M^2/M_p$ (in the case $H \ll  M^2/M_p$ the problem
will be even more pronounced.) In this case the initial energy density on
the wall will be close to its higher-dimensional Planck  value,
$M^{D+4}$. If
one wants to neglect the influence of the bulk on the expansion of the
universe, one should require that the total energy concentrated
there should
be smaller than the energy at the wall. Using Eq. (\ref{masses}), one can
show that this condition  implies that the density of matter in the
bulk at
that time must be smaller than the Planck density $M^{D+4}$ by the
factor of
${M^2/ M_p^2} \ll 1$.
This means that the density of matter in the bulk must be nearly empty as
compared with the density at  the wall, and this emptiness must be
preserved
on a scale $r_0 = M^{-1} \left({M_p\over M}\right)^{2/D} \gg M^{-1}$. In
addition, inflation on the wall requires the distribution  of matter
on the
wall to be homogeneous on scale $\sim H^{-1} \ge {M_p\over  M^2} \gg
M^{-1}$.
Since
$M^{-1}$ is the only natural scale for homogeneity, one can hardly explain
from first principles
how this specific structure could be formed unless there was a
previous stage
of inflation, simultaneously in the bulk and on the wall, which
could extend
the Planck scale $M^{-1}$ to the scale ${M_p/ M^2}$.

Since this subject is rather complicated, we will consider here, for
purely
illustrative purposes, a toy model of the wall inflation. It is  different
from the model of Ref. \cite{savas} but has some obvious similarities.

It is well known that the domain walls in the theories with
spontaneous symmetry breaking in the thin wall approximation can
be described as objects with the energy-momentum tensor $
T^\mu{}_\nu =   \sigma \delta(x) {\rm diag}(1,1,1,0)$, where $x$
is the direction orthogonal to the domain wall, and $\sigma$ is
its surface tension. The metric corresponding to this distribution
of matter describes an inflating $2D$ domain wall in a $4D$
spacetime \cite{vilipsi}\footnote{We would like to extend our
thanks to Raphael Bousso and Ian Kogan, who independently
brought to our attention the possibility to relate domain wall
solutions of \cite{vilipsi} and inflating branes \cite{kogbous}.}.

We have found a similar solution describing a $3D$ inflating domain
wall in a $5D$ space-time.
It is produced by scalar field on the $3D$ domain wall, which in the thin
wall approximation has the energy-momentum tensor
\cite{Solution}
$T^\mu{}_\nu = -\sigma \delta(w) {\rm diag}(1,1,1,1,0)$.
Here $w$ runs along the $5^{th}$ direction, orthogonal to the
3D wall, and $\sigma$ is the surface tension of the 3D wall in the 5D
space-time. We could solve
Einstein's equations with this source to
determine the $3$-brane metric.
It is
\begin{equation}
ds^2_5 = (1 - H  |w|)^2 \Bigl(-dt^2 + e^{2H  t}
(dx^2 + dy^2 + dz^2) \Bigr)
+ dw^2
\label{tribrejn}
\end{equation}
where $H $ is the Hubble constant along the $3$-brane, given by
$H  = {\frac{4\pi \sigma}{3 M^3}}$
\cite{Solution}.  This
metric is perfectly regular at the event horizon
$H  w = 1$, and
singular at the domain wall, where $w=0$. This is the place where the
$\delta$-function
source $T^\mu{}_\nu$ must reside.

The solution (\ref{tribrejn}) can be interpreted as an inflating
$3$-brane, with the event horizon on the brane given by
$H ^{-1}$. In the direction transversal to the brane, the metric
resembles the Rindler metric. The gravitational field is repulsive,
and it pushes the perturbations towards the Rindler horizon, located at
$w = H ^{-1}$.
Any inflating point on the brane is completely surrounded by an
event horizon,
at a distance $H ^{-1}$ from it.
The no-hair theorem
\cite{nht} for this metric would show  that if one perturbs this
metric on a scale greater than $O(H^{-1})$, this perturbation is rapidly
stretched in all directions by the expansion of the universe, just like in
the usual inflationary universe scenario.
For example, if instead of the domain wall at the plane $\omega = 0$ one
considers a spherical domain wall positioned at $x^2+y^2+z^2 + \omega^2 =
r^2$ with $r \gg H^{-1}$ inflation will blow up this bubble, stretch its
walls, and metric near the wall will be again described by our
solution (\ref
{tribrejn}). However, if $r \ll H^{-1}$, then the surface tension of the
domain wall will shrink its size to zero within the time $t \ll
H^{-1}$, and
inflation will never happen. This means, as we expected, that in order to
find out whether inflation on the wall will happen, we need to properly
adjust the initial conditions on a scale $H^{-1}$.

For the scenario the wall inflation in the theory proposed in
Ref. \cite{savas} this would imply that the   wall must be
homogeneous on the
scale $H^{-1} \sim {M^3\over \sigma} \gg M^{-1}$.
This is extremely
difficult to
achieve, especially with $M \sim 1$ TeV, when the size of the initial
homogeneous domain $H^{-1}$ must be 16 orders of magnitude greater
than the
Planck length $M^{-1}$.

In conclusion, we have found that inflation on the wall can be achieved in
several versions
 of inflationary scenario. The simplest way to do so is to use the hybrid
inflation scenario.
 It can be done even for $M \sim 1$ TeV, but it is much easier to do for
larger $M$. However,
 to obtain a complete cosmological scenario one will probably need to
consider not only inflation
  on the wall, but also inflation in the bulk.
  In
such a scenario, since the natural size of homogeneous islands in
the early universe is given by the unification length, $M^{-1} $,
it would be necessary to take the internal dimensions to be
initially small, $\sim M^{-1}$, and allow them to expand until they
reach the
compactification scale $r_0$. During this expansion, the ratio $M /M_p$
may change by many orders of magnitude, and may be much closer to
unity in the
beginning. This may relax the constraints on the mass of the inflaton
\cite{bendav,lyth} quite considerably. The possibility of constructing
bulk inflation could thus play a crucial role in establishing
feasibility of models with large internal dimensions. However,
concrete details of this stage of early inflation may depend on the
specifics
of
the model used to describe it, which lies beyond the scope of the present
article. We will defer this discussion for the future.

Another possible resolution of the outlined problems is related to
the eternal inflation scenario. Indeed, it is known  that the
inflationary universe in simplest versions of chaotic inflation
scenario enters regime of self-reproduction \cite{eternal}. This
means that once inflation begins, it produces infinite amount of
homogeneous space, whereas noninflationary parts of the universe
produce only a finite amount of inhomogeneous space. This fact may
make the problem of initial conditions irrelevant \cite{LLM}. Note
that the regime of self-reproduction occurs not only in chaotic
inflation, but in new inflation as well \cite{Vil}. On the other
hand, this regime does not occur in the pre-big bang inflation,
which makes the problems of initial conditions in this theory more
difficult to resolve \cite{KLB}. If inflation of the wall is
eternal, then we may not necessarily need to have the preceding
stage of inflation in the bulk.

We would like to thank to R. Bousso, I. Kogan and S. Dimopoulos for
helpful
conversations. This work has been supported in part by
NSF grant    PHY-9870115.


\begin{thebibliography}{99}

\bibitem{savas} N. Arkani-Hamed, S. Dimopoulos and G. Dvali,
{\it Phys. Lett.} {\bf B429}, 263 (1998);   hep-ph/9807344;
I. Antoniadis, N. Arkani-Hamed, S. Dimopoulos and G. Dvali,
{\it Phys. Lett.} {\bf B436}, 257 (1998); N. Arkani-Hamed, S.  
Dimopoulos and
J. March-Russell,   hep-th/9809124.

\bibitem{dienes} K. Dienes, E. Dudas
and T. Gherghetta, {\it Phys. Lett.} {\bf B436}, 55 (1998);
  hep-ph/9806293; hep-ph/9807522;
K. Dienes, E. Dudas, T. Gherghetta and
A. Riotto,   hep-ph/9809406.

\bibitem{sundrum} R. Sundrum,   hep-ph/9805471; hep-ph/9807348.

\bibitem{tye} G. Shiu and S.-H. Tye, {\it Phys. Rev.} D {\bf  58}, 106007
(1998);
Z. Kakushadze and S.-H. Tye,   hep-th/9809147.

\bibitem{bachas} C. Bachas, hep-ph/9807415.

\bibitem{benakli} K. Benakli, eprint
hep-ph/9809582.

\bibitem{bendav} K. Benakli and S. Davidson, hep-ph/9810280.

\bibitem{randal} L. Randall and
R. Sundrum,  hep-th/9810155.

\bibitem{witt} P. Ho\v rava and E. Witten, {\it Nucl. Phys.} {\bf  
B460}, 506
(1996);
E. Witten, {\it Nucl. Phys.} {\bf 471}, 135 (1996).

\bibitem{lykken}  J.D. Lykken, {\it Phys. Rev.} D {\bf  54}, 3693 (1996).

\bibitem{antoni} I. Antoniadis and M. Quiros, {\it Phys. Lett.} {\bf  
B392}, 61
(1997).

\bibitem{biq} C.P. Burgess, L.E. Ibanez and
F. Quevedo,   hep-ph/9810535.

\bibitem{rs} V. Rubakov and M. Shaposhnikov,
{\it Phys. Lett.} {\bf B125}, 136 (1983).

\bibitem{ds} G. Dvali and M. Shifman, {\it Nucl. Phys.} {\bf B504},  
127 (1997).

\bibitem{lyth} D. Lyth,   hep-ph/9810320.

\bibitem{andc} A. Linde, {\it Phys. Lett.} {\bf B129}, 177 (1983).

\bibitem{andh} A. Linde, {\it Phys. Lett.} {\bf B259}, 38
(1991); {\it Phys. Rev.} D {\bf 49}, 748 (1994).

\bibitem{term} To avoid terminological misunderstanding we should
emphasize
that hybrid
inflation is a particular version of chaotic inflation. As it was
emphasized
in \cite{andc},
 the  borderline between chaotic inflation and the earlier generation of
inflationary models
 (old inflation and  new inflation) is not in the choice of the
potentials.
The main idea of
 chaotic inflation was that one does not need to assume that the initial
position of the inflaton
 field is fixed by thermal effects; its initial distribution may well be
chaotic. Hybrid inflation scenario \cite{andh} is based on the same idea.


\bibitem{Dterm} P. Binetruy and G. Dvali, {\it Phys. Lett.} {\bf B388},
241 (1996); E. Halyo, {\it Phys. Lett.} {\bf B387}, 43 (1996);
A.D. Linde and A. Riotto,
{\it Phys. Rev.} D {\bf 56}, 1841 (1997); D.H. Lyth and A. Riotto,
hep-ph/9807278

\bibitem{book} A. D. Linde, {\em Particle Physics and Inflationary
Cosmology} (Harwood, Chur, Switzerland, 1990).



\bibitem{nht} G.W. Gibbons and S.W. Hawking,
{\it Phys. Rev.} D {\bf 15} 2738 (1977); S.W. Hawking and I.G. Moss,
{\it Phys. Lett.} {\bf B110}, 35 (1982); W. Boucher and G.W. Gibbons,
in {\it The Very Early Universe}, ed. G.W. Gibbons,
S.W. Hawking and S. Siklos, Cambridge University Press,
Cambridge (1983); A.A. Starobinsky, {\it JETP Lett.}
{\bf 37}, 66 (1983); R. Wald, {\it Phys. Rev.} D {\bf 28} 2118 (1982).

\bibitem{vilipsi} A. Vilenkin, {\it Phys. Lett.} {\bf B133} 177 (1983);
J. Ipser and P. Sikivie, {\it Phys. Rev.}D {\bf 30} 712 (1984).

\bibitem{kogbous} I.I. Kogan, private communication;
R. Bousso, to be published.

\bibitem{Solution}
Our solution (\ref{tribrejn}) can be obtained by directly solving
Einstein's equations in $5D$ with the $3$ brane stress-energy
$T^\mu{}_\nu = - \sigma \delta(w) {\rm diag} (1,1,1,1, 0)$
where $\sigma$ is the generalized surface tension. The $(-)$ sign
is due to our metric signature convention $(-,+,+,+,+)$.
The general form of the metric (\ref{tribrejn}) is an obvious
generalization of
the metric obtained in \cite{vilipsi}). A somewhat nontrivial step
is to find a
relation between $H$ and $\sigma$. To do so, we use the equations of
motion,
and find the Ricci tensor on the brane:
$R^{\alpha}{}_\beta = \frac{8\pi  }{3M^3} \sigma \delta(w)
\delta^\alpha{}_\beta$ where $\alpha, \beta \in \{0,...,3\}$.
Since the curvature is singular,
by symmetry all one needs to check is the divergence of $R^0{}_0$.
The only singularity arises from
$R^0{}_0 (singular) = -\partial_w \Gamma^0_{w 0}= 2 H \delta(w)$.
Hence
$H =\frac{4\pi }{3 M^3} \sigma$,
in contrast to \cite{vilipsi}, where $D=4$ and
$H = {2 \pi  \over  M_p^2}\sigma$.
This completes the derivation of our solution (\ref{tribrejn}).
The solution for a general case of a $D-2$ brane in $D$ dimensions
is very similar, with
$H = \frac{4\pi  }{(D-2) M^{D-2}}\sigma $.
One may also consider a $p$ brane in $D$ dimensions.
The stress energy tensor for it is
$T^\mu{}_\nu = - \sigma \delta(w)
(1_1, ..., 1_{p+1}, 0_{p+2}, ..., 0_D)$,
and the Ricci tensor on the brane is
$R^\alpha{}_\beta = \frac{8\pi (p+3 -D)}{(D-2)M^{D-2}}
\sigma \delta(w) \delta^\alpha{}_\beta$.
For the $3$ brane in $6D$, this gives
$R^\alpha{}_\beta = 0$. In general, for $p+3 \leq D$ one has $R \leq
0$. This
suggests that
the branes in a space with more than one uncompactified dimension
would not
inflate. A particular example is provided by
 cosmic strings in $4D$, for which
$p=1$ and $D=4$. Unlike domain walls, cosmic strings  do not inflate.
This result suggests that in order to have inflation on the wall, we
need to
consider models where
at most one dimension remains
uncompactified. The
class of models proposed in \cite{savas} satisfies this condition.

\bibitem{eternal} A.D. Linde, {\it Phys. Lett.} {\bf B175}, 395 (1986).

\bibitem{LLM} A.D. Linde, D.A. Linde, and  A. Mezhlumian, {\it Phys.  
Rev.} D
{\bf 49},
1783  (1994).

\bibitem{Vil}P.J. Steinhardt, in: {\bf The Very Early Universe}, G.W.
Gibbons, S.W. Hawking, S. Siklos, eds., Cambridge U.P. Cambridge,
England (1982), p. 251;  A.D. Linde,  {``Nonsingular Regenerating
Inflationary Universe,''} Cambridge University preprint (1982);
A. Vilenkin, {\it Phys. Rev.} D {\bf 27}, 2848 (1983).

\bibitem{KLB}  N. Kaloper, A.D. Linde, R. Bousso,  hep-th/9801073.

\end{thebibliography}
\end{document}